\documentclass[conference]{IEEEtran}
\IEEEoverridecommandlockouts

\usepackage{url}
\usepackage{times}
\usepackage{amsmath}
\usepackage{epsfig}
\usepackage{multirow}
\usepackage{longtable}
\usepackage{amsfonts}
\usepackage{amssymb}%
\usepackage{color}
\usepackage{bm}
\usepackage{epstopdf}
\usepackage{subfigure}
\usepackage{cite}
\usepackage{graphicx}
\usepackage{amsmath}
\usepackage{times}
\usepackage{latexsym}
\usepackage[center]{caption2}
\usepackage{stfloats}
\usepackage{cases}
\usepackage{array}
\usepackage{setspace}
\usepackage{fancyhdr}
\usepackage{balance}
\usepackage{algorithm} 
\usepackage{algpseudocode} 

\usepackage[table]{xcolor}
\usepackage{makecell}

        {\par\noindent{\bf Proof.}}%
        {{\hfill{\large$\Box$}}\smallskip}

\begin{document}

\title{Optimal Charging Profile Design for Solar-Powered Sustainable UAV Communication Networks
\thanks{Accepted by IEEE ICC 2023: Green Communication Systems and Networks Symposium.}

}


\author{\IEEEauthorblockN{
Longxin Wang\IEEEauthorrefmark{1},
Saugat Tripathi\IEEEauthorrefmark{2},
Ran Zhang\IEEEauthorrefmark{2},
Nan Cheng\IEEEauthorrefmark{1},
and Miao Wang\IEEEauthorrefmark{2}} \\
\IEEEauthorblockA{\IEEEauthorrefmark{1}
School of Telecommunication Engineering, Xidian University, Xi'an, China}
\IEEEauthorblockA{\IEEEauthorrefmark{2}
Department of Electrical and Computer Engineering, 
Miami University, Oxford, USA}
\IEEEauthorblockA{Email: \IEEEauthorrefmark{1}$wanglx19$@stu.xidian.edu.cn,\IEEEauthorrefmark{1}$dr.nan.cheng$@ieee.org,  \IEEEauthorrefmark{2}\{$tripats,zhangr43,wangm64$\}@miamioh.edu}
}

\maketitle

\begin{abstract}This work studies optimal solar charging for solar-powered self-sustainable UAV communication networks, considering the day-scale time-variability of solar radiation and user service demand. The objective is to optimally trade off between the user coverage performance and the net energy loss of the network by proactively assigning UAVs to serve, charge, or land. Specifically, the studied problem is first formulated into a time-coupled mixed-integer non-convex optimization problem, and further decoupled into two sub-problems for tractability. To solve the challenge caused by time-coupling, deep reinforcement learning (DRL) algorithms are respectively designed for the two sub-problems. Particularly, a relaxation mechanism is put forward to overcome the "dimension curse" incurred by the large discrete action space in the second sub-problem. At last, simulation results demonstrate the efficacy of our designed DRL algorithms in trading off the communication performance against the net energy loss, and the impact of different parameters on the tradeoff performance.

\end{abstract}

\section{Introduction}\label{sec.Intro}

In future mobile communications networks, UAVs equipped with wireless transceivers can be exploited as mobile base stations to provide highly on-demand services to the ground users, forming UAV-based communication networks (UCNs)\cite{fotouhi2019survey}. With the advantages of flexible 3D mobility, higher chance of Line-of-Sight communication channels, and lower deployment and operational cost, UCNs have received substantial research attention from various aspects\cite{mozaffari2019tutorial}. However, the existing works mostly focus on UAV control considering a fixed set of UAVs. Few works have investigated how the network should optimally respond when the UAV crew dynamically change. On one hand, UAVs are powered by batteries. Some UAVs will run out of battery during the service and have to quit the network for charging. On the other hand, supplemental UAVs can be dispatched to easily join the existing crew to enhance the network performance. Therefore, it is indispensable to design novel responsive regulation strategies capable of optimally handling a UAV crew that may change dynamically.

To this end, we proposed in \cite{zhang2020srec,zhang2021learning} a responsive UAV trajectory control strategy to maximize the accumulated number of served users over a time horizon where at least one UAV quits or joins the network. Nevertheless, no matter how good such responsive strategies can be, they are by nature passive reaction strategies which can only accept and passively react to the change rather than proactively control the change. Solar charging makes the proactive control possible. The chance lies in that the user traffic demand in an area is usually time-varying. When the demand is low, some UAVs can be deliberately dispatched to elevate high to get solar charged even if they are not in bad need of charging. They can be called back later to replace other UAVs or meet the increased user demand. In this manner, the network is able to take charge of the change in the serving UAV crew, and a solar-powered sustainable (SPS) UAV network can be established.

UAV communications with solar charging have been studied by some pioneering works. With a single solar-powered UAV, \cite{sun2019optimal} developed an optimal 3D trajectory control and resource allocation strategy, \cite{sekander2020statistical} studied the problems of energy outage at UAV and service outage at users by modeling solar and wind energy harvesting, and \cite{zhang2020power} proposed a novel power cognition scheme to intelligently adjust the energy harvesting, information transmission, and trajectory to improve UAV communication performance. For multi-UAV networks, the work\cite{khairy2020constrained} studied joint dynamic UAV altitude control and multi-cell wireless channel access management to optimally balance between solar charging and communication throughput improvement. The work\cite{turgut2020energy} analytically characterized the user coverage performance of a UAV network based on a harvest power model and 3D antenna radiation patterns. Although solar charging is exquisitely integrated to fuel the UAV communications, most of the related works do not take into account the time-variability of solar radiation or user traffic demand, which, however, is usually the case in practice. To achieve day-scale sustainability, it is essential to consider these time variation when designing the UAV control strategies.

Therefore, in this work, we investigate the optimal solar charging strategy design for a UCN, considering time-varying solar radiation and user data traffic demand. The strategy aims to optimally trade off over a time horizon between maximizing the accumulated user coverage and minimizing the net energy loss, subject to the constraints of UAV sustainability and user service requirements. The net energy loss is defined as the difference between the total energy harvest and the total energy consumption. As far as we know, this work is the first to jointly consider time-varying solar radiation and user service demand at a day-scale in a solar-powered self-sustainable UAV network. Specifically, our contributions are three-fold.
\begin{itemize}
    \item The studied problem is first formulated into a time-coupled mixed-integer non-convex optimization problem. To make the problem tractable, the original problem is decoupled into two sub-problems, one obtaining the mapping between the number of serving UAVs and the number of served users in each time slot, and the other handling the time-variability of solar radiation and user service demand.
    \item To tackle the challenge caused by time coupling, deep reinforcement learning (DRL) algorithms are developed to solve the two sub-problems. Particularly, a relaxation mechanism is designed to relieve the ``dimension curse" caused by the large discrete action space in the second sub-problem.
    \item Simulations are conducted to demonstrate the efficacy of the proposed learning algorithms and the impact of different parameters on the tradeoff performance.
\end{itemize}

remainder of the paper is organized as follows. Section II depicts the system model. Section III presents the problem formulation and decomposition. Section IV details the proposed DRL algorithms and its relaxation. Section V provides the numerical results. Section VI concludes the paper.
\section{System Model}\label{sec.SystemModel}
\begin{figure}[!ht]
	\centering
	\includegraphics[width=3.4in]{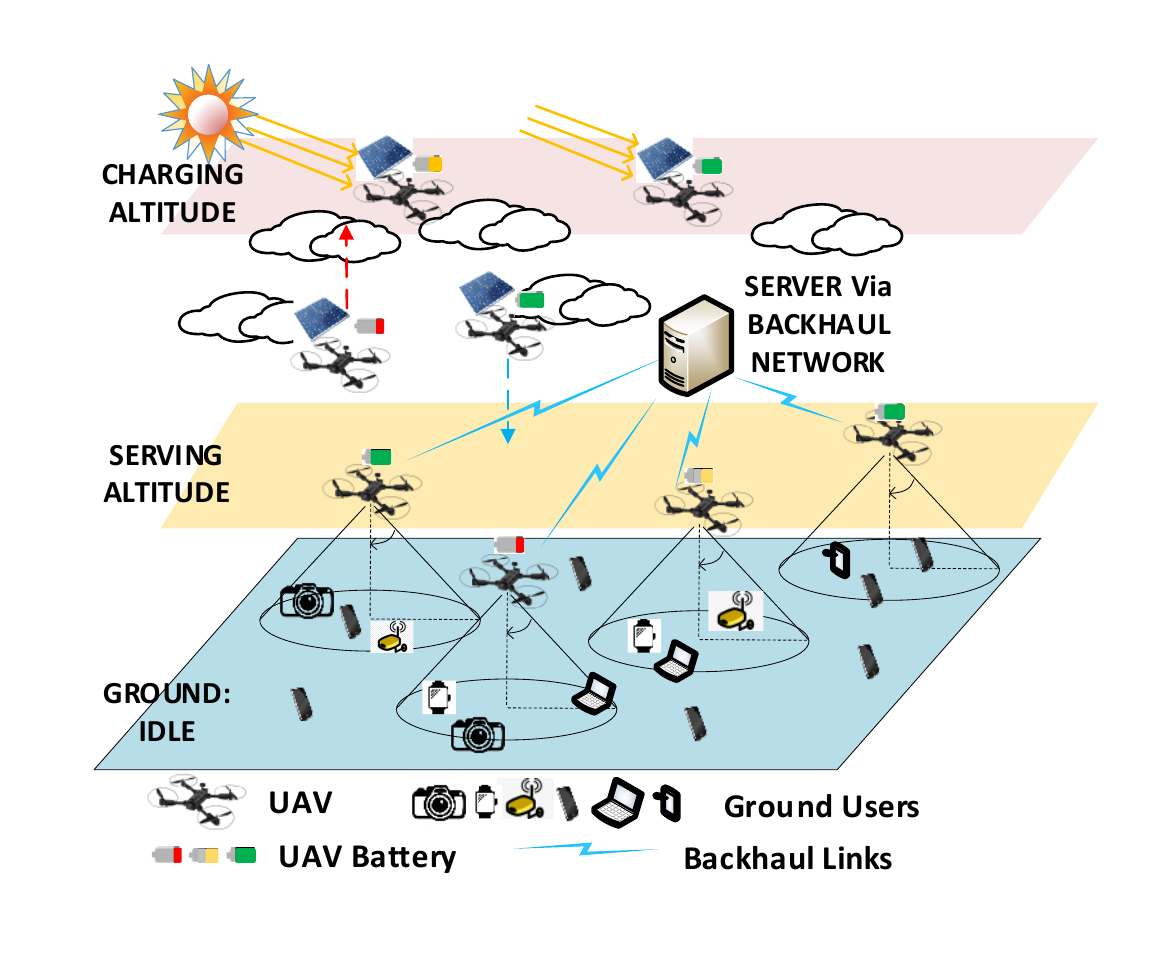}
	\caption{System model of SPS UAV communication network.} \label{fig.SystemModel}
\end{figure}
\subsection{Network Model}\label{subsec.NetMdl}
As shown in Fig. \ref{fig.SystemModel}, we consider $N$ solar chargeable UAVs, denoted as $\mathcal{S_{UAV}}$, providing communication services to (i.e., serve) a target area. All UAVs can communicate with a server via backhaul networks (e.g., a satellite or cellular network). 
Each UAV concentrates its transmission energy within an aperture underneath it, forming a ground coverage disk. UAVs are mostly at three altitudes: ground, the serving altitude ($H_{Srv}$) and the charging altitude ($H_{Chg}$). When on the ground, UAVs consume negligible power only for messaging with the server. UAVs only serve and get charged at the fixed altitude $H_{Srv}$ and $H_{Chg}$, respectively. $H_{Srv}$ is low to maintain good UAV-user communication quality, while $H_{Chg}$ is right above the upper boundary of the clouds to minimize the attenuation of solar radiation due to clouds. The consideration of only charging UAVs at $H_{Chg}$ is justified as follows. According to \cite{kokhanovsky2004optical}, the solar radiation is attenuated exponentially with the thickness of clouds between the sun and solar panel, leading to only $\sim\frac{1}{10}$ after the first 300 meters. When it does not take long (e.g., one minute or two) for a UAV to move vertically through 300 meters, UAVs can be reasonably set for charging at a fixed altitude just above the clouds.

The time horizon $T$ is equally divided into time slots indexed by $t$. In any time slot, a percentage $p$ of the users are randomly distributed in proximity to some hotspot centers while the remaining are uniformly distributed throughout the area. The numbers and spatial distributions of the users and hotspots are deemed unchanged within a time slot but may vary with $t$. The dynamics of the user distribution are known to the server to obtain an offline UAV charging strategy. The well-trained strategy will be executed relying on the server-UAV communications via the backhaul links. 

\subsection{Spectrum Access}\label{subsec.sa}
Users access the UAV spectrum following LTE Orthogonal Frequency Division Multiple Access (OFDMA)\cite{zhang2015probabilistic}, which assigns different users of one UAV at least one orthogonal Resource blocks (RBs) such that they do not interfere with each other. A heuristic 2-stage user association policy is adopted. In each time slot, users send connection requests in stage I to the serving UAVs that provides the best SINR (can be measured via reference signaling), and each UAV admits the users with the best SINR values based on its bandwidth. In stage II, the rejected users then chooses the UAV with the next best SINR and is admitted if the UAV has available bandwidth. The stage II procedure is repeated for each user without association until it is admitted or has no available UAV to send requests to. Each user has a minimum throughput requirement $r_u$. When a UAV $i$ admits a user $u$, the number of RBs assigned to the user, $n^{RB}_{iu}$, should satisfy

\begin{equation}\label{eq.sinr}
\begin{array}{l}
    n^{RB}_{iu}W^{RB}\log_2(1+\frac{P_tG_{iu}}{n_0+\sum_{j\in {\mathbf{S}_u}'\backslash\{i\}}P_tG_{ju}})\ge r_u\\
    \text{where }G_{iu}=10^{-PL_{iu}/20},\\
    \;\;\;\;\;\;\;\;\;\;PL_{iu} = 20\log_{10}{(\frac{4\pi f_cd_{iu}}{c})}+\eta \;\;\; \text{(dB)}.
\end{array}
\end{equation}

In Eq. \eqref{eq.sinr}, $W^{RB}$ is the bandwidth per RB, $P_t$ is the transmit power spectrum density (psd) of UAVs, $n_0$ is the noise psd, ${\mathbf{S}_u}'$ denotes the set of UAVs that can cover user $u$, and $G_{iu}$ is the UAV-to-user channel gain as a function of the center frequency $f_c$, distance $d_{iu}$ between UAV $i$ and user $u$, light speed $c$ and a line of sight (LoS) related parameter $\eta$\cite{al2014modeling}.

\subsection{Energy Model}\label{subsec.EMdl}
We follow the work in \cite{sun2019optimal} to model the kinematics power consumption for the UAVs. For a UAV flying at a level speed $v_{lv}$ and a vertical speed $v_{vt}$, the kinematics power consumption is modeled as Eq. \eqref{eq.pc}. In Eq. \eqref{eq.pc}, $P_{lv}$, $P_vt$ and $P_{drag}$ denotes the power consumption of level flight, vertical flight and blade drag profile power, respectively, $W$ is the UAV weight, $\rho$ is the air density, $A$ is the total area of the UAV rotor disks, $C_{D0}$ is the profile drag coefficient, $\sigma A$ is the total blade area, and $v_T$ is the blade tip speed. Note that $v_{vt}$ is positive for UAV climbing, and negative for UAV landing.
\begin{equation}\label{eq.pc}
\begin{array}{l}
    P_{kine} = P_{lv} + P_{vt} + P_{drag},\\
    \text{where }P_{lv} =\frac{W^2}{\sqrt{2}\rho A}\frac{1}{\sqrt{v_{lv}^2+\sqrt{v_{lv}^4+4V_h^4}}},\\
    \;\;\;\;\;\;\;\;\;\;P_{vt} = Wv_{vt},\\
    \;\;\;\;\;\;\;\;\;\;P_{drag} = \frac{1}{8}C_{D0}\rho \cdot\sigma A||v_T||^3\\
    \;\;\;\;\;\;\;\;\;\;V_h = \sqrt{\frac{W}{2\rho A}}.
\end{array}
\end{equation}

In addition to the kinematics power consumption, UAVs spend power on communication and on-board operations like computing, which are denoted as $P_{tx}$ and $P_{static}$, respectively. Thus the total power consumption of a UAV is given as
\begin{equation}\label{eq.tot_pc}
P_{Tot} = P_{kine}+P_{tx}+P_{static}.
\end{equation}
Note that $P_{kine}$ is usually several hundred watts. The transmission power of a small BS covering hundreds of meters typically falls between 0.25$W$ and $6W$\cite{txpower}. The operational power consumption is also in single-digit watts. Thus, $P_{tx}$ and $P_{static}$ are usually neglected in practice. 

The solar radiation intensity above the clouds varies with time in a day. We follow the model in \cite{sekander2020statistical} to characterize the average intensity as
\begin{equation}\label{eq.solarintense}
    I_{rad}(t) = \max\{0, I_{max}(-1/36t^2 + 2/3t -3)\}, 0\le t< 24,
\end{equation}
where $t$ represents hour $t$, and $I_{max}$ denotes the maximum intensity during a day. The harvested solar power is then calculated as
\begin{equation}\label{eq.harvest}
    P_{h}(t) = \left\{
    \begin{array}{l}
         A_{c}\frac{\eta_c}{K_c}I_{rad}(t)^2, \;\;\;\;0<I_{rad}(t)<K_c,\\
         A_{c}\eta_c I_{rad}(t) \;\;\;\;\;\;\;I_{rad}(t)\ge K_c,
    \end{array}\right.
\end{equation}
where $A_{c}$ is the solar panel area, $\eta_c$ is the charging efficiency coefficient, and $K_c$ is an intensity threshold.
\section{Problem Formulation and Decomposition}\label{sec.PF}
The objective is to achieve the optimal tradeoff among maximizing the total number of served users over the time horizon $T$, maximizing the total harvested solar energy, and minimizing the total energy consumption of the UAV network. The optimization is subject to the network sustainability constraints and user traffic demand requirements. With the above considerations, the problem formulation is given as $P_1$.

In Problem $P_1$, the decision variables include whether a UAV should land, go serving or go charging at any time slot $t$, i.e, $\hat{\mathbf{a}}_{t}=(a_{1,t},\cdots,a_{N,t})$, and the horizontal positions of the UAVs that go serving in any time slot $t$, i.e., $\hat{\mathbf{p}}_{t}=(a_{k_1(t),t},\cdots,a_{k_M(t),t})$, where $k_m(t),m\in\{1,\cdots,M\}$ index the serving UAVs at time slot $t$. Part $A_1$ denotes the amount of energy harvest from solar charging for UAV $i$ at time slot $t$. This part is determined by $a_{i,t}$ and $a_{i,t-1}$ since a UAV takes some time to move from the last altitude to the current one, the harvest solar power $P_h(t)$, and the battery residue at the end of time slot $t-1$, i.e., $E^{res}_{t-1}$, since the battery capacity may be reached during charging. Part $A_2$ represents the energy consumption of UAV $i$ at $t$, which is also determined by $a_{i,t}$, $a_{i,t-1}$, and $E^{res}_{t-1}$. In Part $A_3$, $\hat{\mathcal{S}}^u_t(\cdot)$ is the set of users that are admitted and served by all the UAVs at $t$, which is a function of $\hat{\mathbf{a}}_{t}$ and $\hat{\mathbf{p}}_{t}$. The constant $C$ is the coefficient balancing the weights between the user coverage and the energy gain and losses.
\begin{alignat}{2}
\max\limits_{\hat{\mathbf{a}}_{t},\;\hat{\mathbf{p}}_{t}}&\quad\sum\limits^{T}_{t=1}\left\{C\sum\limits_{i\in \mathcal{S_{UAV}}}[\underbrace{E_{h}(a_{i,t},a_{i,t-1},P_{h}(t),E^{res}_{i,t-1})}_{A_1}\right.\tag{$P_1$}\label{P1}\\
&\quad\left.-\underbrace{E_{c}(a_{i,t},a_{i,t-1},E^{res}_{i,t-1})}_{A_2}] + \underbrace{|\hat{\mathcal{S}}^u_t(\hat{\mathbf{a}}_t,\hat{\mathbf{p}}_t)|}_{A_3}\right\}\nonumber\\
\mbox{s.t.} & \quad E^{res}_{i,t}\ge E_{min}(a_{i,t}),\;\forall i \mbox{ and }\forall t;\tag{C1.1}\\
&\quad|\hat{\mathcal{S}}^u_t(\hat{\mathbf{a}}_t,\hat{\mathbf{p}}_t)|\ge p_{min}|\mathcal{S}^u_t|,\;\forall t;\tag{C1.2}\\
&\quad \mbox{Eq.\eqref{eq.sinr}},\;\forall u\in\hat{\mathcal{S}}^u_t(\hat{\mathbf{a}}_t,\hat{\mathbf{p}}_t)\mbox{ and }\forall t.   \tag{C1.3}
\end{alignat}

Constraint C1.1 represents the network sustainability requirements. The battery residue of any UAV $i$ at any $t$ should be no smaller than an altitude-dependent threshold $E_{min}(a_{i,t})$. This is to make sure at the end of each $t$, each UAV has enough energy to elevate to $H_{Chg}$ for charging in future slots to avoid completely leaving the crew. Thus,
\begin{equation}\label{eq.minE}
    E_{min}(a_{i,t}) = \Delta H/v_{up}\cdot({P_{Tot}}_{|v_{lv}=0,v_{vt}=v_{up}}),
\end{equation}
where $\Delta H$ takes $H_{Chg}$ when $a_{i,t}$=0, $H_{Chg}$-$H_{Srv}$ when $a_{i,t}$=1, and 0 when $a_{i,t}$=2. Constraints C1.2 and C1.3 represent the user data traffic demand requirements. C1.2 requires that the percentage of served users at any $t$ should be no less than $p_{min}$ given the total number of users $|\mathcal{S}^u_t|$. C1.3 requires that the individual user traffic demand $r_u$ should be satisfied for any served users at any $t$.

Problem $P_1$ is a mixed integer nonlinear non-convex optimization problem with nonlinear constraints. The items of different $t$ in the objective function are temporally coupled through UAV battery residue. These facts make the sequential decision problem intractable. To this end, we decouple $P_1$ into two sub-problems $P_2$ and $P_3$, each of which can be solved by means of DRL algorithms. 
\begin{alignat}{2}
\max\limits_{\hat{\mathbf{p}}_{t}}&\quad|\hat{\mathcal{S}}^u_t(\hat{\mathbf{p}}_t,N^{Srv}_{UAV})|,\quad\forall t\tag{$P_2$}\\
&\quad \mbox{Eq.\eqref{eq.sinr}},\;\forall u\in\hat{\mathcal{S}}^u_t(\hat{\mathbf{p}}_t,N^{Srv}_{UAV}).   \tag{C2.1}
\end{alignat}
In the first sub-problem $P_2$, given the user distribution at each $t$ and the number of UAVs in service $N^{Srv}_{UAV}$, the total number of users that can be served is maximized via determining the optimal positions $\hat{\mathbf{p}}^*_t(N^{Srv}_{UAV})$ as a function of $N^{Srv}_{UAV}$ and the user distribution. 

\begin{alignat}{2}
\max\limits_{\hat{\mathbf{a}}_{t}}&\quad\sum\limits^{T}_{t=1}\left\{C\sum\limits_{i\in \mathcal{S_{UAV}}}[E_{h}(a_{i,t},a_{i,t-1},P_{h}(t),E^{res}_{i,t-1})\right.\tag{$P_3$}\label{P3}\\
&\quad\left.-E_{c}(a_{i,t},a_{i,t-1},E^{res}_{i,t-1})] + |\hat{\mathcal{S}}^u_t(\hat{\mathbf{a}}_t,\hat{\mathbf{p}}^*_t(N^{Srv}_{UAV}))|\right\}\nonumber\\
\mbox{s.t.} & \quad E^{res}_{i,t}\ge E_{min}(a_{i,t}),\;\forall i \mbox{ and }\forall t;\tag{C3.1}\\
&\quad|\hat{\mathcal{S}}^u_t(\hat{\mathbf{a}}_t,\hat{\mathbf{p}}^*_t(N^{Srv}_{UAV}))|\ge p_{min}|\mathcal{S}^u_t|,\;\forall t;\tag{C3.2}\\
&\quad N^{Srv}_{UAV}=\sum\nolimits_{i\in\mathcal{S_{UAV}}}I(a_{i,t}==1),\;\forall t\label{eq.relation}.
\end{alignat}
In the second sub-problem $P_3$, the achieved mapping from P2 between the maximum number of served users $|\hat{\mathcal{S}}^u_t(\hat{\mathbf{a}}_t,\hat{\mathbf{p}}^*_t(N^{Srv}_{UAV}))|$ and $N_{UAV}^{Srv}$ is exploited. The same objective as in $P_1$ is maximized via only optimizing $\hat{\mathbf{a}}_{t}$. The relationship between $N^{Srv}_{UAV}$ and $\hat{\mathbf{a}}_{t}$ is given in Eq. \eqref{eq.relation}, where $I(\cdot)$ is a binary indicator taking 1 if the inside condition is true and 0 otherwise.

\section{Design of Deep Reinforcement Learning Algorithms}\label{sec.Algorithm}
In this section, design of the DRL algorithms for solving $P_2$ and $P_3$ are elaborated. For $P_2$, we reuse the algorithm designed in our previous work \cite{zhang2021learning} to achieve the mapping between the number of UAVs in service ($N_{Srv}^{UAV}$) and the total number of served users ($|\hat{\mathcal{S}}^u_t|$) given a certain user distribution in hour $t$. Based on the time-dependent mappings, the DRL algorithm design for solving $P_3$ is emphatically presented.

\subsection{DRL for Solving $P_2$}
Our previous work \cite{zhang2021learning} considered a set of UAVs flying at a fixed height providing communication services to the ground users with minimum throughput requirement. We considered dynamic UAV crew change during the training due to battery depletion or supplementary UAV join-in. A DDPG algorithm was designed to maximize the user satisfaction score via obtaining the optimal UAV trajectories during the steady period without crew change and the transition period when crew changes. With the following simplifications, the proposed algorithm can be fit to $P_2$. Crew change is not considered so that the state space can be cut down to only including the UAV positions. The action space remains unchanged allowing a UAV to go any direction with a maximum distance $d_{max}$ per step. The reward function changes from step-wise user satisfaction score to step-wise total number of served users. The closest-SINR based user association policy is replaced with that elaborated in Subsection \ref{subsec.sa}.

\subsection{DRL for Solving $P_3$}
P3 exploits the mappings between $|\hat{\mathcal{S}}^u_t|$ and $N_{Srv}^{UAV}$ in different hours obtained via $P_2$, and aims to maximize its objective function via optimizing the UAVs' charging profiles in the considered time horizon. In each hour $t$, the DRL agent needs to determine whether a UAV should go charging, go serving or go to the ground for energy saving, i.e., $\hat{\mathbf{a}}_{t}$, based on the UAVs' current battery residues and altitudes, solar radiation intensity, and user traffic demands. When designing the DRL algorithm, we consider the varying solar radiation and user traffic demands as the dynamics of the underlying environment. The key components of the algorithm are designed as follows.

\subsubsection{State Space} Battery residue of a UAV is a critical factor in determining its next move, thus $\{E^{res}_{i,t}\}, \forall i\in\mathcal{S_{UAV}}$ are included into the state space, denoting the residue battery of UAV $i$ at the beginning of hour $t$. The current UAV altitude is another in-negligible factor as the altitude changing will incur wear energy consumption which may accumulate to significantly impact the overall scheduling. Minimizing the unnecessary altitude changes for UAVs while satisfying the constraints will contribute positively to the optimization objective. Therefore, $\{H_{i,t}\},\forall i\in\mathcal{S_{UAV}}$ are embraced in the state space, which takes values $H_{Chg}$, $H_{Srv}$, or 0. Lastly, the hour indexing $t$ needs to be considered to capture the dynamics of the environment (e.g., solar radiation and user traffic demand) so that different actions may be taken at different $t$ even if the rest of the states are same. The complete state space is given below, with a cardinality of $2N+1$.
\begin{equation}
    \mathbf{S_t} = \{E^{res}_{i,t},H_{i,t},t\},~~~~\forall i\in\mathcal{S_{UAV}}.
\end{equation}

\subsubsection{Action Space} The decision variables of subproblem $P_3$ is $\hat{\mathbf{a}}_{t}=(a_{1,t},\cdots,a_{N,t})$, denoting the altitudes that each UAV will go to at the beginning of the current hour $t$. The action space is directly defined as $\mathbf{A_t}=\{a_{i,t}\},\forall i\in\mathcal{S_{UAV}}$, which takes value 0 if the UAV goes to the ground, 1 if the UAV goes serving, and 2 if the UAV goes charging. The cardinality of the action space is $3^N$.

\subsubsection{Reward Function Design} The reward function $r_t$ consists of three parts. The first part $r_{1,t}$ corresponds to the constraints of $P_3$. When any UAV breaks the sustainability constraint (C3.1), a constant penalty $p_{C_1}<0$ is applied. When constraint C3.2 is broken, i.e., the total number of serving UAVs $N^{Srv}_{UAV}$ cannot result in a minimum user service rate $p_{min}$, a constant penalty $p_{C_2}\in(p_{C_1},0)$ is applied. In addition, when $N^{Srv}_{UAV}$ is larger than the minimum number of serving UAVs that result in 100$\%$ user service rate, a reward of 0 is applied to prevent service over-provisioning.

The second part $r_{2,t}$ corresponds to the maximization of the total number of served users over the entire time horizon. Thus, $r_{2,t}$ is set to be directly equal to $|\hat{\mathcal{S}}^u_t(\hat{\mathbf{a}}_t,\hat{\mathbf{p}}^*_t(N^{Srv}_{UAV}))|$. The third part $r_{3,t}$ corresponds to the maximization of the difference between the total harvested energy and the total consumed energy. Due to the time-varying solar radiation intensity, it is beneficial for a UAV to land to the ground if it does not serve during some hours of a day (e.g., in the night or around sunset/sunrise), while it is beneficial to go charging during other hours of a day. In the former case, positive reward is given for UAVs going to the ground to promote energy saving, whereas in the latter case, positive reward is given for UAVs going charging to encourage energy harvesting. Therefore, $r_3$ is designed as
\begin{equation}
    r_{3,t} = \left\{\begin{array}{l}
    c_1\cdot N^{Gnd}_{UAV},\text{ if landing is beneficial at }t;\\
    c_2\cdot N^{Chg}_{UAV},\text{ if charging is beneficial at }t,
    \end{array}\right.
\end{equation}
where $c_1$ and $c_2$ are reward coefficients to tradeoff between $A_1$ and $A_3-A_2$ in $P_1$, replacing coefficient $C$ for $A_1$. The total instantaneous reward $r_t$ is $r_t=r_{1,t}+r_{2,t}+r_{3,t}$.

\subsubsection{Relaxation of the Discrete Action Space} As the state space $\mathbf{S_t}$ is continuous and discrete mixed, and the action space $\mathbf{A_t}$ is discrete, Deep Q learning (DQL) algorithm is typically exploited. However, the cardinality of the action space, i.e., $3^N$, increases exponentially with the total number of UAVs $N$. For an instance of $N$=15, the total number of possible aggregate actions over all UAVs will be $3^{15}\approx$1.4$e^7$. As the number of outputs of the Deep Q network (DQN) is equal to the total number of possible actions, the resultant DQN will be prohibitively complicated, not to mention the number of hours in the considered time horizon. Therefore, DQL is technically feasible, but practically impossible. 

Inspired by \cite{dulac2015deep}, we relax the original discrete action space into a continuous space and obtain the UAV charging profile $\hat{\mathbf{a}}_{t},\forall t\in T$ by means of DDPG. Each action $a_{i,t}$ is relaxed from discrete values $\{0,1,2\}$ to a continuous range (-0.5, 2.5). Hence, the relaxed action space becomes $\mathbf{\widetilde{A_t}}=\{a_{i,t}\}\in(-1.5,2.5)^N$. With DDPG, the number of outputs of the actor network is equal to the dimension of the action space, i.e., $N$, which only increases linearly with $N$ rather than exponentially as $3^N$ in DQN. Each time when an aggregate action is determined by the actor network and added with noise, the action will then be discretized to the closest value in $\{0,1,2\}$. The discretized action will be the actual action applied to the current state and stored in the experience replay buffer. In this manner, the complexity of the exploration can be significantly reduced.

\section{Numerical Results}\label{sec.Simulation}

\subsection{Simulation Setup}
For subproblem $P_2$, we reuse the simulation setup and parameter configurations in our previous work \cite{zhang2021learning} to obtain the hourly mapping between the number of UAVs and the maximal number of served users. For subproblem $P_3$, the environment parameters and the RL parameters are summarized in Table \ref{Table:env} and \ref{Table:RL}, respectively. A 24-hour time horizon is considered per episode with each hour being a step. The training is conducted using Reinforcement Learning Toolbox of Matlab 2022a on a Windows 10 server with Intel Core i9-10920X CPU @ 3.50GHz, 64GB RAM, and Quadro RTX 6000 GPU. 

Note that only considering 24 hours will not guarantee the same set of UAVs working for consecutive days, but only ensure that the involved UAVs have sufficient battery residue to go charging in the next day. Once it is verified that a given set of UAVs can be sustainable for a whole day, two sets of UAVs can serve in alternative days to achieve full sustainability.

\begin{table}[!ht]
\footnotesize
\centering
\renewcommand{\arraystretch}{1}

\begin{tabular}{!{\vrule width0.8pt}l|l!{\vrule width0.8pt}}\Xhline{0.8bp}
\multicolumn{1}{!{\vrule width0.8pt}c|}{\gape{\bfseries Parameters}} & \multicolumn{1}{c!{\vrule width0.8pt}}{\gape{\bfseries Values}} \\ 



  \hline
  \rowcolor[gray]{0.9}
  UAV levels $(H_g,H_s,H_c)$ & (0,300,1400)$m$\\
  Max. solar radiation intensity  & $2000W/m^2$\\
  above clouds $I_{max}$ & \\
  \rowcolor[gray]{0.9}
  Solar radiation intensity threshold $K_c$ & $150W/m^2$\\
  UAV charging efficiency $\eta_c$ & 0.25\\
  \rowcolor[gray]{0.9}
  UAV maximum speeds $(v_{lv},v_{up},v_{dn})$ & ($6m/s$, 4$m/s$, 4$m/s$)\\
  UAV rotor disk radius $r_d$ & 0.3$m$\\
  \rowcolor[gray]{0.9}
  UAV charging panel area $A_{Chg}$ & $1m^2$\\
  UAV weight and air density $(W,\rho)$ & (5x9.8$N$, 1.225$kg/m^3$)\\
  \rowcolor[gray]{0.9}
  UAV Battery Capacity $E_{cap}$ & $600 Wh$\cite{Bat_Cap}\\
  UAV static operational power $P_{static}$ & $5W$ \\
  \rowcolor[gray]{0.9}
  UAV drag profile $(C_{D0},\sigma, v_T)$ & $(5e$-$4,0.056,150m/s)$\\
  
  Minimum user service rate $p_{min}$ & 0.85\\
 
 

  \hline
  \end{tabular}
\caption{Summary of Main Environment Parameters}\label{Table:env}
\end{table}

\begin{table}[!ht]
\footnotesize
\centering
\renewcommand{\arraystretch}{1}

\begin{tabular}{!{\vrule width0.8pt}l|l!{\vrule width0.8pt}}\Xhline{0.8bp}
\multicolumn{1}{!{\vrule width0.8pt}c|}{\gape{\bfseries Parameters}} & \multicolumn{1}{c!{\vrule width0.8pt}}{\gape{\bfseries Values}} \\ 



  \hline
  \rowcolor[gray]{0.9}
  Actor and critic networks & 2 hidden layers, each with\\
  \rowcolor[gray]{0.9}
  & $400$ hidden nodes\\
  
  Neural network learning rates & $0.0001$ for critic and actor\\
  \rowcolor[gray]{0.9}
  Activation function & $tanh$ (actor output), \\
  \rowcolor[gray]{0.9}
  & $Relu$ (all remaining)\\
  
  Regularization & L2 with $\lambda=0.0001$\\
  \rowcolor[gray]{0.9}
  Gradient threshold & 1\\
  
  Smooth factor for target networks & 0.001\\
  \rowcolor[gray]{0.9}
  Update frequency for target networks & 1\\
  
  Mini-Batch size & 512\\
  \rowcolor[gray]{0.9}\rowcolor[gray]{0.9}
  Action noise ($\sigma^2_{max}$, decay,$\sigma^2_{min}$) & $(1.5, 0.0001, 0.2)$\\
  
  Experience buffer capacity & $10^6$ \\
  \rowcolor[gray]{0.9}
  Discount factor $\gamma$ & $0.99$ \\
  
  Max. steps per episode & $24$\\
  \rowcolor[gray]{0.9}
  Max. episodes simulated & $10^5$\\

  \hline
  \end{tabular}
\caption{Reinforcement Learning Parameters}\label{Table:RL}
\end{table}

\subsection{Simulation Results}\label{subsec.simuresults}
\begin{figure}[!ht]
\begin{tabular}{c}
\subfigure [Solar radiation dynamics]{\includegraphics[height=1.1in, width=3.3in]{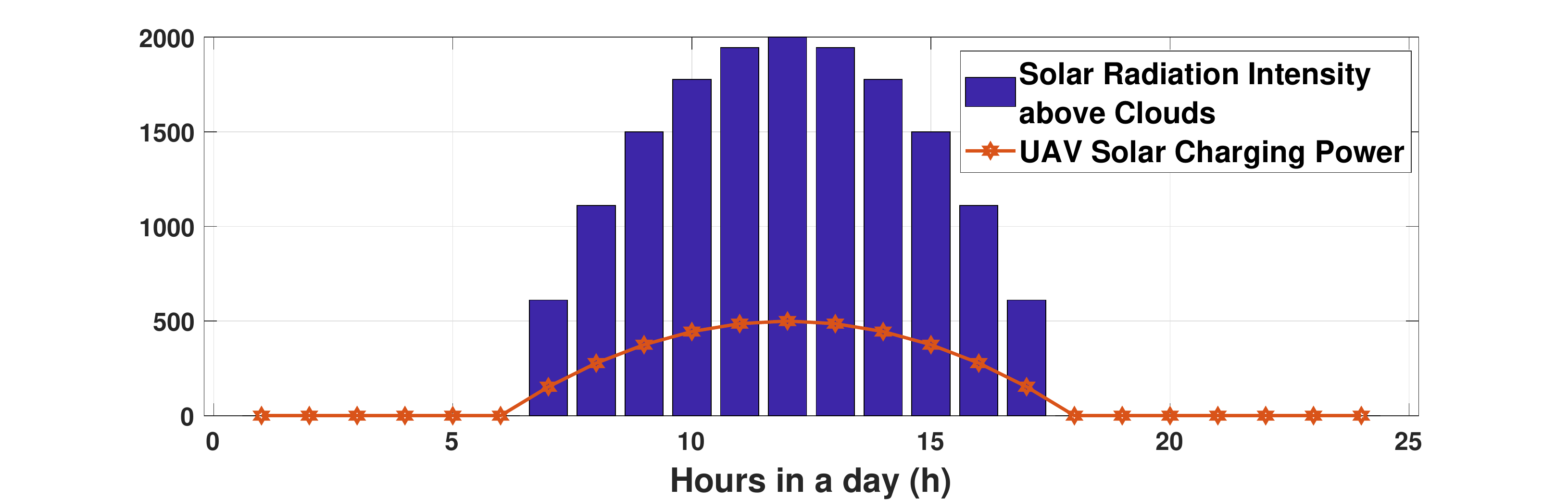}\label{Fig.Dynamics1}}\\
\subfigure [User demand dynamics]{\includegraphics[height=1.1in]{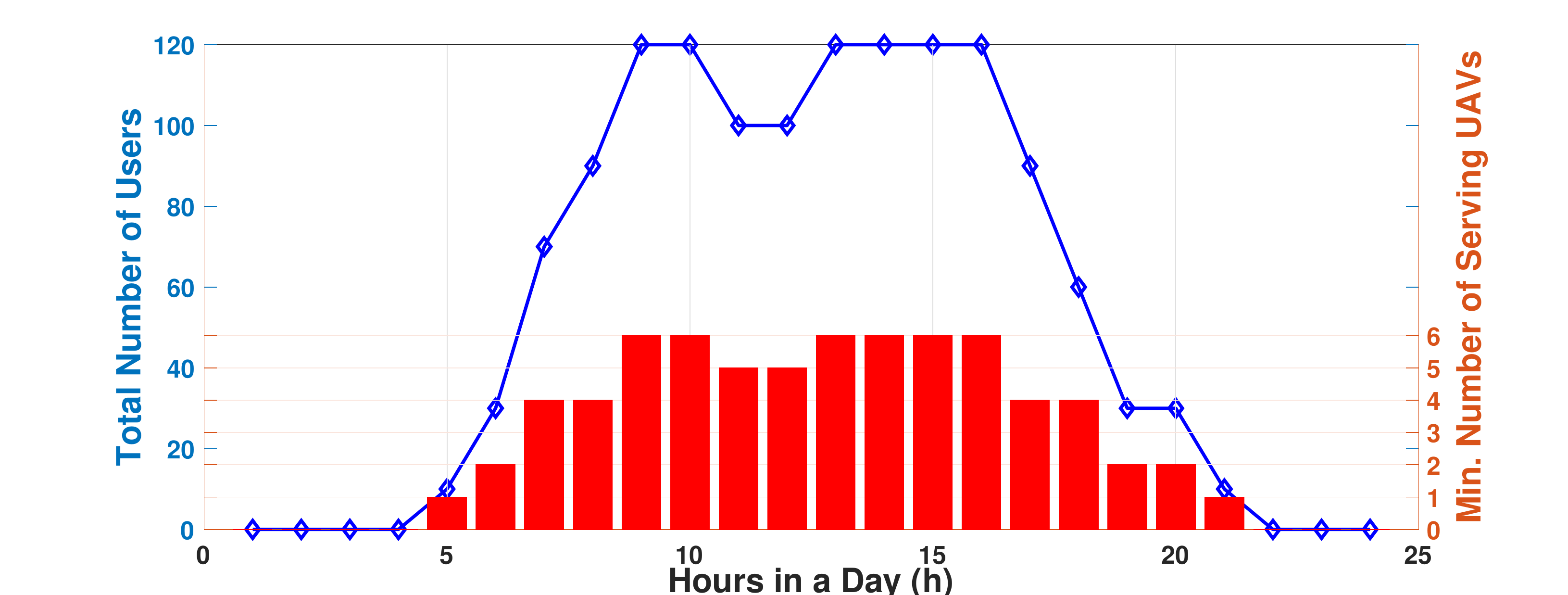}\label{Fig.Dynamics2}} 
\end{tabular}
\caption{\small{Dynamics of Solar Radiation and User Demand in a Day.}}\label{fig.Dynamics}
\end{figure}

The major dynamics of the environment are presented in Fig. \ref{fig.Dynamics} first. The solar radiation is concentrated between 7am-5pm in a day. The UAV charging rate above clouds is positively related to the solar radiation. In Subfig. \ref{Fig.Dynamics2}, there are more users requesting services in the late morning and afternoon, which is consistent with daily working hours. To satisfy a minimum $85\%$ of user service rate per hour, the red bars show the minimum number of UAVs needed for serving.

\begin{figure}[!ht]
	\centering
	\includegraphics[width=3.2in]{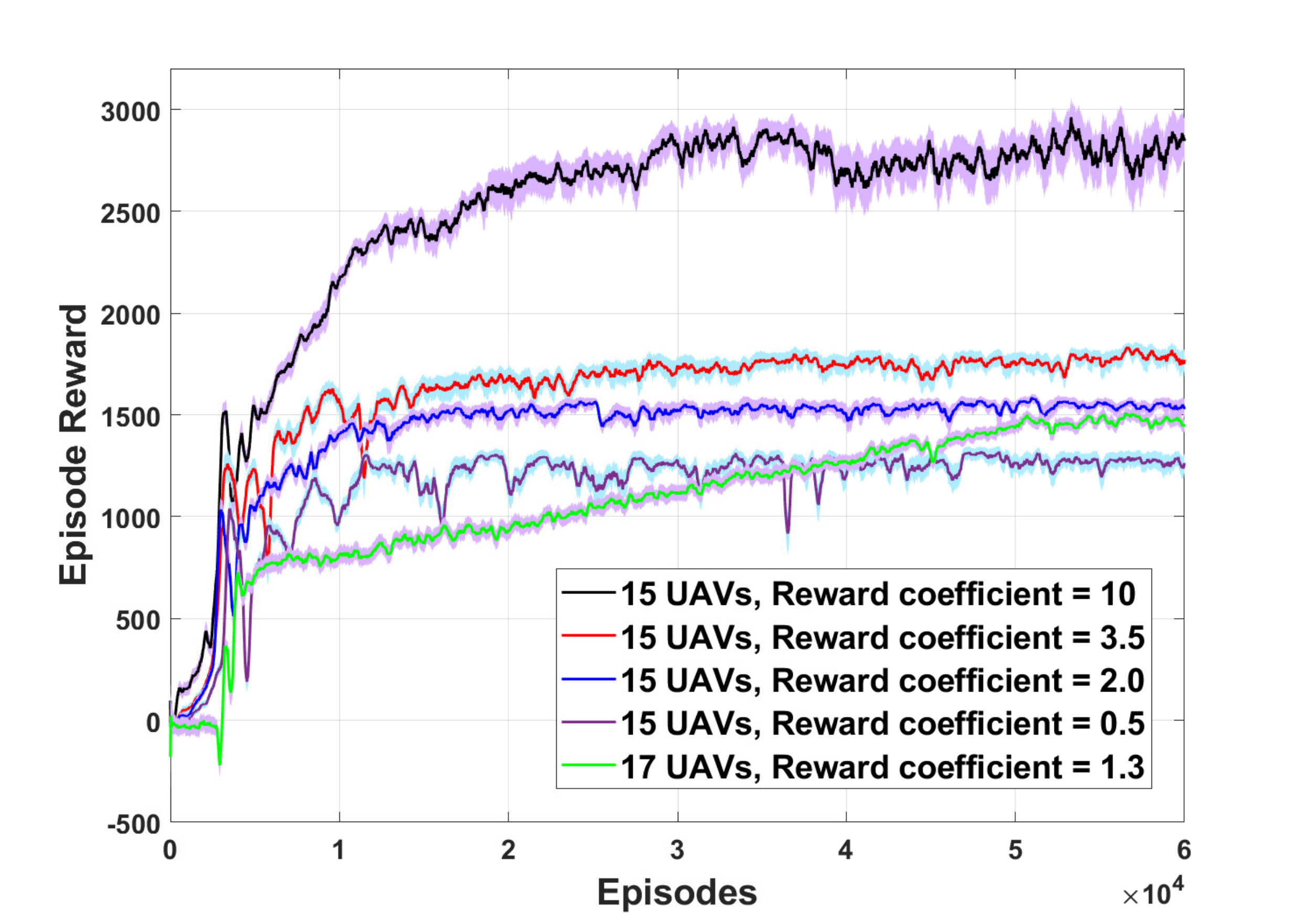}
	\caption{\small{Convergence of episodic reward for different numbers of UAVs and reward coefficients ($c_1,c_2$). The episode rewards are averaged over a window size of 300 with 95$\%$ credit interval.}} \label{fig.Converge}
\end{figure}

Given the above dynamics and the constraints of sustainability and user demand, the convergence of episodic rewards under our designed DRL algorithms are provided in Fig. \ref{fig.Converge}. It can be seen that for the same number of UAVs, convergence is almost the same for different reward coefficients, yet a larger UAV set will lead to longer convergence time due to increased dimensions of the state-action space.

\begin{figure}[!ht]
\begin{tabular}{c}
\subfigure [Hourly number of serving UAVs]{\includegraphics[height=1.1in, width=3.3in]{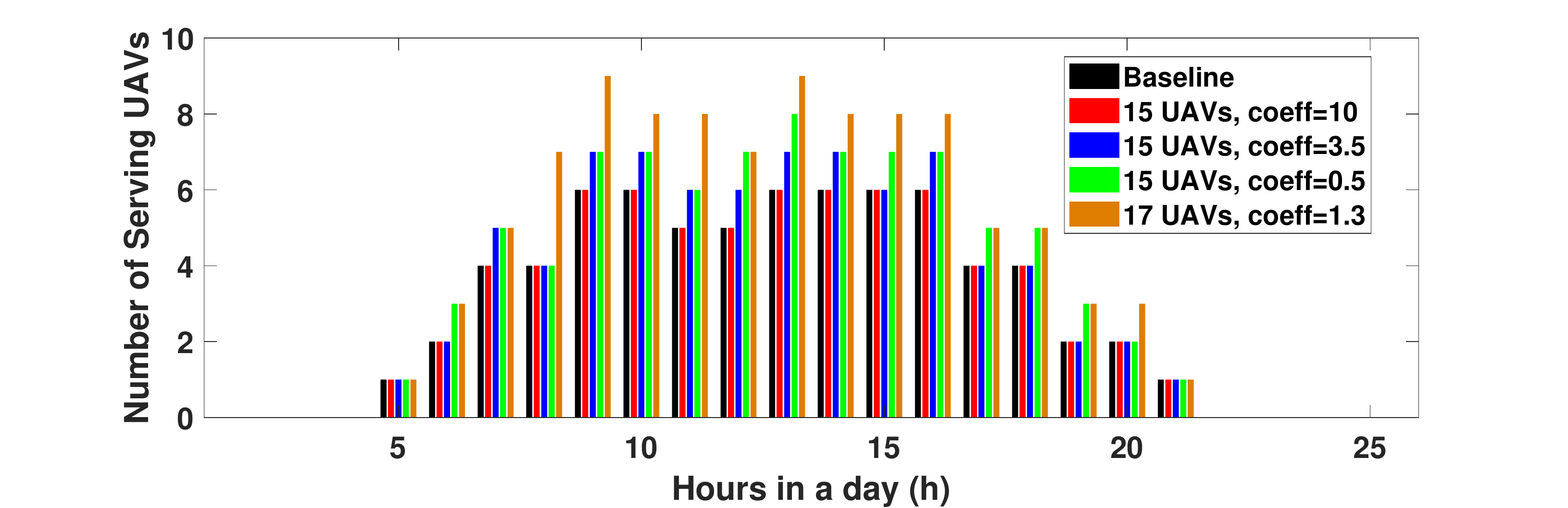}\label{Fig.Episode1}}\\
\subfigure [Accumulated number of served users]{\includegraphics[height=1.1in, width=3.3in]{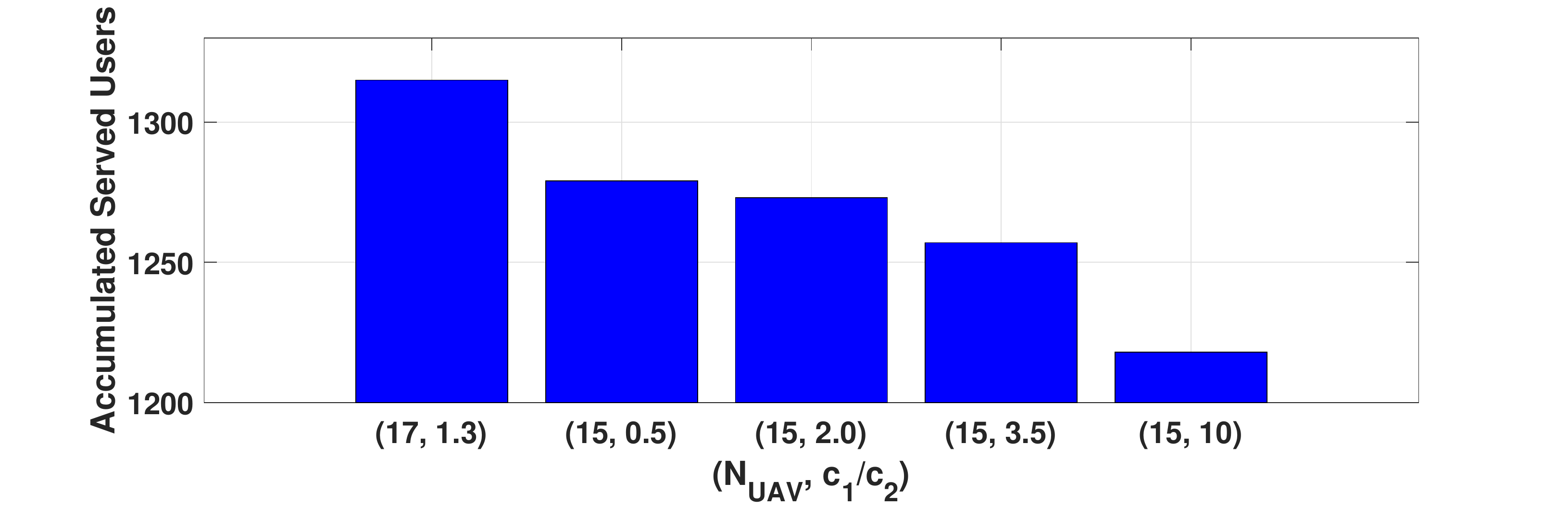}\label{Fig.Episode2}} 
\end{tabular}
\caption{\small{Performance measures of the proposed algorithm under different parameters.}}\label{fig.Details}
\end{figure}

Fig. \ref{fig.Details} reveals details of the achieved optimal charging profiles in terms of hourly number of serving UAVs and the accumulated number of served users in one day. The baseline in Subfig. \ref{Fig.Episode1} gives the minimum required number of serving UAVs in each hour to satisfy the $85\%$ user service rate. It can be observed that with smaller reward coefficients ($c_1$, $c_2$), the number of serving UAVs in each hour tend to increase. The reason is that smaller reward coefficients will result in a smaller weight $C$ in optimization $P_1$. Thus the RL agent tends to dispatch more UAVs to serve more users to get more rewards rather than make UAVs go charging or idle. When there are more UAVs available (e.g., 17 UAVs), more UAVs can serve in each hour when $c_1$ and $c_2$ are relatively low. More serving UAVs in each hour will consequently bring more served users, which is confirmed by Subfig. \ref{Fig.Episode2}.

\section{Conclusions}\label{sec.Conclusion}
In this paper, optimal solar charging problem has been studied in a sustainable UAV communication network, considering the dynamic solar radiation and user service demand. The problem has been formulated into a time-coupled optimization problem and further decoupled into two sub-problems. DRL algorithms have been designed to make the sub-problems tractable. Simulation results have demonstrated the efficacy of the designed algorithms in optimally trading off the communication performance against the net energy loss.

\bibliographystyle{IEEEtran}
\bibliography{main}

\end{document}